\documentclass[
    ,final            
  ]
  {aipproc}

\layoutstyle{6x9}

\newcommand{\degr}{\hbox{$^\circ$}}

\begin{document}

\title{First Results From Sleuth:\\ The Palomar Planet Finder}

\author{Francis T. O'Donovan}{
  address={California Institute of Technology, M/C 105-24, Pasadena,
  CA 91125},
}

\author{David Charbonneau}{
  address={California Institute of Technology, M/C 105-24, Pasadena,
  CA 91125},
}

\author{Lewis Kotredes}{
  address={California Institute of Technology, M/C 105-24, Pasadena,
  CA 91125},
}

\begin{abstract}
We discuss preliminary results from our first search campaign for
transiting planets performed using Sleuth, an automated 10 cm
telescope with a 6 degree square field of view. We monitored a field
in Hercules for 40 clear nights between UT 2003 May 10 and July 01,
and obtained an rms precision (per 15-min average) over the entire
data set of better than 1\% on the brightest 2026 stars, and better
than 1.5\% on the brightest 3865 stars. We identified no strong
candidates in the Hercules field. We conducted a blind test of our
ability to recover transiting systems by injecting signals into our
data and measuring the recovery rate as a function of transit depth
and orbital period. About 85\% of transit signals with a depth of 0.02
mag were recovered. However, only 50\% of transit signals with a depth
of 0.01 mag were recovered. We expect that the number of stars for
which we can search for transiting planets will increase substantially
for our current field in Andromeda, due to the lower Galactic latitude
of the field.
\end{abstract}

\maketitle


\section{Acquisition and Analysis of Sleuth Observations}

Sleuth\footnote{\url{http://www.astro.caltech.edu/~ftod/sleuth.html}}, located at Palomar Observatory in Southern California, is the
third transit-search telescope in our network which comprises
STARE\footnote{\url{http://www.hao.ucar.edu/public/research/stare/stare.html}}
(PI: T. Brown, located in Tenerife), and PSST (PI: E. Dunham, located
in northern Arizona). Sleuth is an f/2.8 lens with a 10 cm aperture
that images a $6\degr\times6\degr$ field of view onto a $2048 \times
2048$ back-illuminated CCD camera. Sleuth conducts nightly
observations with an SDSS r' filter, but also gathers color images in
g', i' \& z' during new moon. Sleuth automatically adjusts the focus
for changes in temperature and filter. A separate f/6.3 lens feeds the
guide camera. The automated observations, including operation of
the clamshell enclosure, are controlled by a workstation running
Linux. In the event of threatening weather, the on-site night
assistant for the 200" telescope can close the system remotely, and an
observatory weather station provides additional protection. At dawn,
the night's data are automatically compressed and sent by ftp to our
workstation at Caltech.

\begin{figure}[tb]
\includegraphics[width=7.5cm]{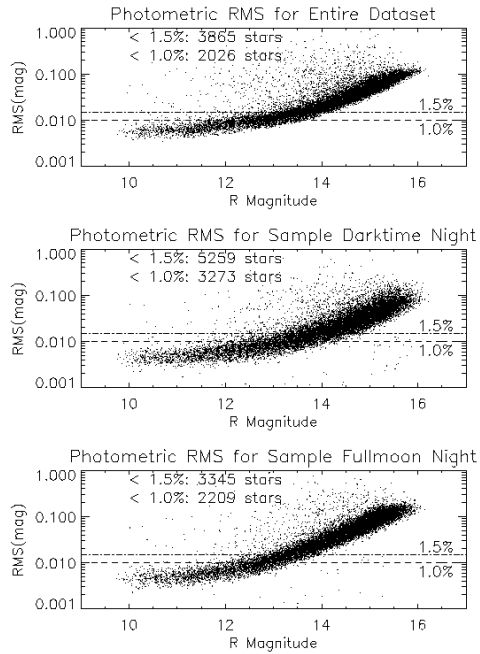}
\caption{The calculated rms as a function of R magnitude for the
10,000 brightest stars in our Hercules field for three subsets of
data: (upper panel) all observations; (center panel) observations from
the night of June 29th UT during new moon; (lower panel) observations
from the night of June 14th UT during full moon. The numbers of stars
with rms below 1.0\% and 1.5\% for each dataset are shown.}
\label{fig:rms_plot}
\end{figure}

Between UT 2003 May 10 and July 01, we monitored approximately 10,000
stars (\mbox{$9 < R < 16$}) in a field in
Hercules. Figure~\ref{fig:rms_plot} shows the calculated rms error in
our photometry. We applied the STARE photometry code (written by
T. Brown [HAO/NCAR] and with adaptations by G. Mandushev [Lowell
Obs.]) to calibrate and perform weighted-aperture photometry upon the
images. We subsequently combined the time series for each star for all
nights. These light curves were then processed through a cleaning
pipeline to remove discrepant data and correlated variations in
magnitude. Finally, the data were averaged in 15 minute bins to
produce final light curves.

\begin{figure}[tb]
\includegraphics[width=5.5cm,angle=90]{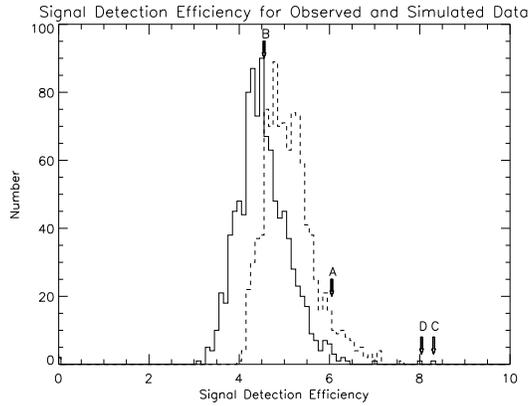}
\caption{ The solid line shows the histogram of derived values for the
Signal Detection Efficiency (SDE) for the best-fit transit signal for
each of the 1000 brightest stars. The histogram of derived SDE values
for our simulated data set is plotted as the dashed line. The phased
time series for two stars with large SDE values (indicated as C and D
on the plot) are shown in Figure~\ref{fig:all3_plot}. We also inserted
transits into the data and attempted to recover these: phased light
curves of two such recovered systems are shown in
Figure~\ref{fig:AandB_plot}, and the resulting SDE values are
indicated on the plot for the actual data (A) and the simulated data
(B).}
\label{fig:hist_plot}
\end{figure}


We then used the box-fitting algorithm of
\citet{KovacsZuckerMazeh:2002} to search these light curves for
transits with periods ranging from 1.5 to 7.5 days. This program
assigned a Signal Detection Efficiency (SDE -- see
\citep{KovacsZuckerMazeh:2002}) to the star, based on the significance
of the transit detection. Figure~\ref{fig:hist_plot} shows a histogram
of the SDEs of the thousand brightest stars in the Hercules field, and
compares it with the SDEs of a thousand simulated light curves with a
Gaussian noise distribution of the same mean and variance as the
original timeseries.

\begin{figure}[tb]
\includegraphics[width=7.5cm]{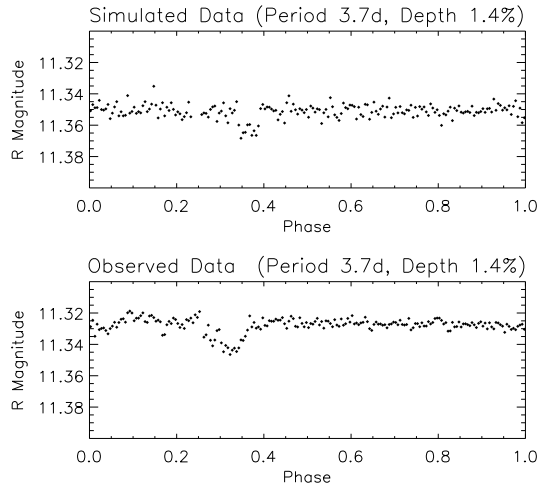}
\caption{As a test of our ability to recover transit signals, we
injected transits of varying period and depth into the data, as well
as into simulated data. The recovered 1.4\%-deep transits for a 3.7
day period for two such stars are shown for the simulated data (upper
panel) and actual data (lower panel). The SDE values for these
recoveries are shown in Figure~\ref{fig:hist_plot}.}
\label{fig:AandB_plot}
\end{figure}

We also tested the transit search code by inserting simulated transits
into our photometric data and attempting to recover these signals -- an
example of such an injected transit is shown in
Figure~\ref{fig:AandB_plot}. The recovery rate of these transits is
about 85\% for transits with a depth of 0.02 mag, and about 50\% for
0.01 mag transits. 

\section{The Next Steps}

Throughout September and October 2003, we gathered photometric data
on a field in Andromeda.  The lower Galactic latitude of this field
($b=-16\degr$ versus $b=+40\degr$ for the Hercules field) will lead
to a significantly larger number of stars that we can search for
transiting planets. The Andromeda data also benefit from a firmware
upgrade to our guider, which removed the large night-to-night pixel
offsets that the Hercules data were subject to. Whereas the typical
offsets in Hercules were 20 pixels, in our current field these offsets
rarely exceed 5 pixels, and guiding within a single night is often
good to 3 pixels.

We will perfom more detailed tests of our transit detection algorithm
to calibrate and improve the detection probability. We are concerned
that the cleaning pipeline itself may be introducing photometric
correlations between the target stars, and thus reducing our detection
capability. We intend to introduce simulated transits into the data
before cleaning, and verify that this processing does not remove these
events.

\begin{figure}[tb]

\includegraphics[width=7.5cm]{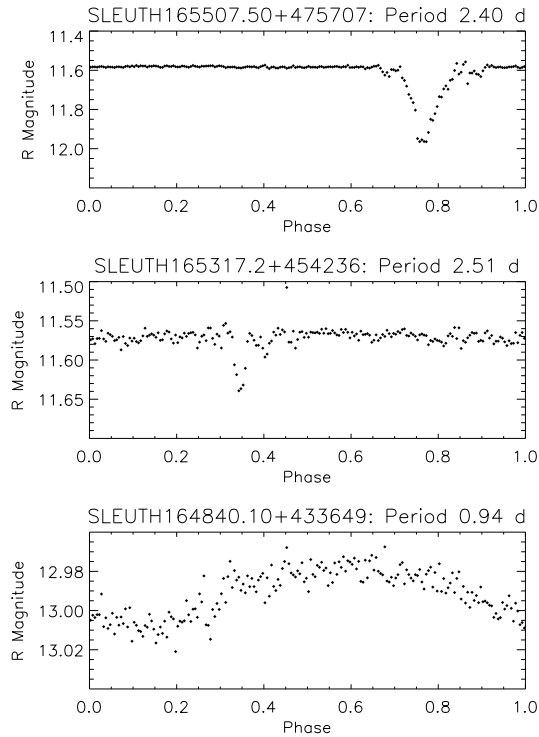}
\caption{ The phased light curve for: (C -- upper panel) a
grazing-incidence eclipsing binary system, and (D -- center panel) a
low-amplitude eclipsing system. The SDE values for these systems are
shown in Figure~\ref{fig:hist_plot}. The phased light curve of a
detected low-amplitude variable stars is also shown (lower panel).}
\label{fig:all3_plot}

\end{figure}

Example Sleuth light curves of eclipsing binaries are shown in
Figure~\ref{fig:all3_plot}. Figure~\ref{fig:all3_plot} also shows the
light curve of a variable star. We intend to compile a catalog of such
binaries and variables. A dominant concern for any transit survey is
the rejection of false positives, i.e. systems containing an eclipsing
binary that mimic the photometric light curve of a transiting
planet. (The eclipse depths in Figure~\ref{fig:all3_plot} are too
great to be mistaken for planetary transits.) To minimize
resource-intensive spectroscopic follow-up work, we are planning to
conduct our own high angular-resolution, multi-color photometry of
transit candidates. To this end, we are currently building Sherlock
(\citet{Kotredes_et_al.:SFOW}), an automated follow-up telescope for
wide-field transit searches.

\bibliography{aamnemonic,mybib}
\bibliographystyle{aipproc}   

\end{document}